\documentstyle [12pt] {article}

\begin{document}

%-----------------------------------------------------------
% The Document
%-----------------------------------------------------------

\title {Theory and Phenomenology of Mixed Amphiphilic Aggregates}
\author {Michael M. Kozlov$^{1}$ and David Andelman$^2$ \\
\\
         $^1$Department of Physiology and Pharmacology\\
         Sackler School of Medicine\\
         Fax: +972-3-6409113; Phone: +972-3-6407863\\
         E-Mail: misha@devil.tau.ac.il\\
and\\
         $^2$School of Physics and Astronomy   \\
         Raymond and Beverly Sackler Faculty of Exact Sciences \\
         Fax: +972-3-6422979; Phone: +972-3-6407239 \\
         E-Mail: andelman@post.tau.ac.il\\
\\
         Tel Aviv University, Ramat Aviv, Tel Aviv 69978, Israel \\
         \\}
\date{May 1996}
\maketitle

\noindent

\vspace{1cm}
\begin{abstract}
\setlength {\baselineskip} {20pt}

We give a short overview of existing approaches describing
shapes and energetics
of amphiphilic aggregates. In particular, we consider recent
experimental data and theory in relation to {\em mixed} aggregates.
We point out the outstanding questions deserving further
investigations such as stability of single-component vesicles and
size growth of mixed vesicles
induced by micelle-forming surfactants.
\end{abstract}

\pagebreak

\setlength{\baselineskip}{20pt}
\setcounter{equation}{0}

\section {Introduction}
An amphiphilic
molecule consists of two moieties with opposing properties:
hydrophilic polar head and hydrophobic hydrocarbon tail. Such
hydrophilic-hydrophobic nature of amphiphiles determines their
 self-assembly and fascinating physical properties.
Numerous investigations
have recently addressed
structures and phase behavior of amphiphiles in aqueous solutions
\cite{luzatti}.
This interest is partially motivated by the
significant role played by
amphiphiles in biophysics, biotechnology and
pharmacology \cite{lasic,barenholz}.
Indeed, an important class of amphiphiles, called
phospholipids, form the basis of cell membranes, while various
surfactants such
as octyl glucoside  (OG) and bile salts are used for extraction
of proteins from cell membranes (membrane solubilization) and
for preparation of artificial phospholipid membranes containing specific
proteins (membrane reconstitution).

Self-assembly in
water has many features in common for different amphiphiles.
First,
as the amphiphile concentration becomes higher than a critical one,
 the amphiphilic molecules
assemble into  monomolecular layers (monolayers), which are about
$\delta\simeq 1.5$nm thick and
have one surface covered by the hydrophilic heads and the second formed by
the hydrophobic tails. Then, these monolayers
self-organize into structures
where all the hydrophobic surfaces are shielded
from contact with water by the hydrophilic ones.

The difference between aggregates formed by various amphiphiles is in
the shapes of the monolayers
\cite{luzatti,lichtenberg,rand,seddon,koynova}.
We describe the shape of a monolayer by its
characteristic curvature $c$ or radius of curvature $R =1/ c$. A
more exact definition will be given below. The monolayers of many common
phospholipids such as lecithin have nearly flat geometries
so that their radii
of curvature are much larger than the monolayer thickness, $R>>\delta$.
In water they form flat bilayers organized
as multi-lamellae  or unilamellar vesicles
\cite{lasic}.
The latter (so-called liposomes for
phospholipids) are small ``bags" made of amphiphilic bilayers
of dimensions in the range of few nm to
several $\mu$m. Another type of amphiphiles,
called detergents (micelle-forming surfactants),
forms in many cases monolayers which are
strongly curved away from the aqueous surrounding
 with a radius
of curvature of the order of the
monolayer thickness, $R\simeq\delta$. Due
to their high curvature such monolayers exist in water in the form of
cylindrical or spherical micelles.

Besides flat bilayers and micelles,
which are the main subject of this presentation, few other
characteristic shapes are observed for amphiphile monolayers. One of them
is also characterized by fairly high curvatures
opposite to the curvature of micelle-forming
surfactants
(directed away from
the hydrophobic tails).
Such amphiphiles (for example, dioleoylphosphatidylethanolamine (DOPE))
form in water an
inverted hexagonal phase (so-called the H$_{II}$)
at high enough concentrations \cite{rand,koynova}.
Surfactants of another type ({\em e.g.},
monoolein) tend to form monolayers of saddle-like shape. The
corresponding structures in water are referred to as bicontinuous  cubic
phases \cite{seddon}.

While the shapes of monolayers formed by individual
amphiphiles are known in
many cases, they are less well understood in the case
of monolayers consisting of mixtures of different
amphiphiles. In particular,
 mixtures of  bilayer-forming lipids
and micelle-forming surfactants are of importance
 to membrane solubilization and reconstitution, as has been recently
 investigated by several experimental
groups. Since detailed reviews of this phenomenology exist
\cite{lichtenberg,rigaud},
we give here only a brief summary of the
results.

\begin{itemize}
\item
 The unilamellar liposomes in a single-component lipid (lecithin)
retain their shape and size for long times following
their preparation.

\item
Upon addition of small amounts of micelle-forming
surfactant to the lipid vesicles,
the system stays in the form of stable mixed vesicles. However, the
vesicle size changes as a function of  concentration of the
added surfactant. In  most cases vesicle growth is observed and
cannot be explained only by the addition of the surfactant material.
The increase of the vesicle size can be associated
with processes such as fusion and material transport mediated
by the micelles.
However, in recent studies of solubilization of vesicles
consisting of dimeric amphiphiles \cite{danino}, an opposite effect of
a decrease
in the vesicle size has been observed by cryo-transmission
electron microscopy (cryo-TEM).

\item
When the amount of surfactant in the system reaches
 a critical surfactant-to-lipid ratio $R_e^{SAT}$,
a first-order phase transition from the
mixed liposomes into mixed micelles
is observed.
%At a higher value of the surfactant-to-lipid ratio called
%$R_e^{SOL}$ all the vesicles have been solubilized into mixed micelles.

The resulting micelles have, in most cases,
a thread-like shape. However,
direct formation of spheroidal micelles (and lack of
thread-like micelles)
has been reported
for the so-called dimeric or gemini surfactants,
({\em e.g.,}
alkanediyl-$\alpha,\gamma$-bis (dimethyldodecylammonium bromide)
\cite{danino}. In studies of solubilization of lecithin vesicles by
a cationic surfactant cetyltrimethylammonium chloride (CTAC)
a new structure intermediate between liposomes and thread-like
micelles has been reported \cite{edwards}. This intermediate
structure consists of
perforated bilayers {\em i.e.},  liposomes filled with holes.

The transition of liposomes into micelles progresses with increase
of surfactant
to lipid ratio $R_e$ from $R_e^{SAT}$ to a higher value
called $R_e^{SOL}$.
Inside this range the
mixed liposomes seem to be in thermodynamical equilibrium with mixed
micelles.

\item
At $R_e^{SOL}$ only micelles are present in the aqueous solution.
Further increase of the amount of surfactant results in the case of
thread-like micelles in reduction of their sizes till they become
spheroidal.

\end{itemize}

Variety of shapes of amphiphilic
monolayers and the transitions between them
pose a challenge to theoreticians.
Two  theoretical approaches describing such systems should be
mentioned. In one 
detailed molecular
structure and the microscopic interactions 
are taken into account \cite{blankschtein}. 
In this review  we describe a different 
approach to surfactant systems in terms of
elastic properties of monolayers and try to
point out the questions deserving further analysis.

\section{Models of Membrane Elasticity}

We mention several basic models describing elastic
properties of amphiphilic
systems. The first formulated in 1973 by Helfrich
\cite{helfrich} determines elastic
behavior of lipid membranes whose shapes only slightly deviate from a
flat
surface. The local shape of the membrane surface can be characterized
by its
principal curvatures $c_1$ and $c_2$, and its elastic energy
per unit area is written
as

\begin{equation}
f = {1\over 2}\kappa (c_1 + c_2 - c_0)^2 + \bar\kappa c_1 c_2
\label{1}
\end{equation}
where the elastic properties of the flat membrane are determined by the
bending modulus $\kappa$, the Gaussian curvature modulus
$\bar\kappa$ and the spontaneous curvature $c_0$. While $\kappa$ has a
common meaning of rigidity with respect to changes of the curvature,
$c_0$ and $\bar\kappa$ have a more delicate physical origin.
The spontaneous
curvature $c_0$ accounts for non-vanishing stress in the flat
membrane
($c_1 = c_2 = 0$). In qualitative terms $c_0$ reflects a tendency
of the flat
membrane to curve spontaneously in order to relax its internal stresses.
The modulus of the Gaussian curvature $\bar\kappa$ has a
physical meaning of membrane stress related to the Gaussian curvature
$c_1\cdot c_2$.
The energy determined by $\bar\kappa$ depends, according to the
Gauss-Bonnet theorem, only on the membrane topology, and changes at
processes such as fusion or fission of liposomes. The basic
assumption in (1) is that the radii of curvature are much bigger
than the membrane thickness, $|c_1\delta|$, $|c_2\delta| \ll 1$,
so that only contributions up to second order in these
small parameters are retained.

The Helfrich model (1)
turned out to be very useful in analyzing the elastic
and thermodynamical properties of lipid vesicles. However,
the description of micelles by (1) is
not straightforward because of the high
curvatures existing in the micellar packing.

Another approach describing formation of amphiphilic
aggregates in terms of
effective molecular dimensions has been proposed by Israelachvili et al
\cite{israelachvili}.
It makes use of the effective packing parameter $p$ of an amphiphilic
molecule, which combines the molecular volume $v$, the molecular area $a$ and
the length of hydrocarbon chain in its fully extended state $l$ such that
$p = v / (a l)$. In crude terms, formation of micelles is
expected for  $p < 1$, molecules with $p \simeq 1$ are
supposed to form flat monolayers (bilayers in water),
while amphiphiles with $p > 1$
are predicted to form inverted cylinders of the H$_{II}$ phase.
Although this model explains qualitatively shapes of amphiphilic
aggregates, it
does not allow to calculate the energy of monolayers and the
 phase transitions between  different structures.

An elastic model of strongly curved monolayers has been formulated by
Gruner \cite{gruner} for the inverted monolayers of
H$_{II}$ phases and
has been later extended for monolayers of arbitrary shapes
\cite{kozlov1}. Formally, Gruner model is similar to
the Helfrich one, as
the energy per unit area of a cylindrical monolayer is related to its
curvature by

\begin{equation}
  \label{2}
f = {1\over 2}\kappa (c - c_{in})^2
\end{equation}
where $\kappa$ is the bending modulus. The parameter $c_{in}$ called
the intrinsic curvature is formally analogous to the Helfrich spontaneous
curvature $c_0$. However, it has a different physical meaning. While
$c_{0}$ accounts for the bending stress in the flat state
of the monolayer,
$c_{in}$ determines the monolayer shape in a completely
relaxed stress-free
state. The model (2) assumes that the curvature only
slightly deviates from
its intrinsic value so that $| (c - c_{in})/ c_{in}|\ll 1$,
and the bending energy is expanded up to
quadratic order
in this small value.
On the other hand and in contrast to model (1), the curvature
itself can be arbitrarily high.

The model (2) in its
extended form \cite{kozlov2} has been recently used to
explain experimental results on osmotic-stress deformations of the
strongly curved cylindrical monolayers of H$_{II}$ phases and phase
transitions between these monolayers and the flat bilayers of lamellar
phases. In those studies it has been assumed that the quadratic expression
(2) is valid also for strong bending deformations where
$|(c - c_{in})/ c_{in}|\simeq 1$. In other words, it was assumed
that the bending
stress-strain relationships are linear in a wide range
of deformations. The good agreement between the predictions of the model
and the experimental results,
indeed, supports this assumption.
On the other hand, the physical consequences
of
this result have
still to be understood in terms of microscopical structure
of monolayers.

The elastic moduli $\kappa$, $\bar\kappa$ and the spontaneous curvature
$c_0$ of monolayers introduced phenomenologically in the pioneering theory
(1) have been computed in  detail by a mean-field approach developed in
the series of works and has been reviewed recently
\cite{benshaul}. The analysis of the monolayer material parameters
in terms of the  hydrocarbon chain and
polar head molecular structure
  contributes greatly to our understanding of monolayer
elasticity at a microscopic level.

\section{Elastic Parameters of Mixed Monolayers}

The existing theoretical approaches describing monolayers composed
of different kinds of amphiphile molecules are based on the original
theories (1) and (2), where the elastic moduli $\kappa$, $\bar\kappa$ and
the spontaneous curvature
$c_0$ are determined by the elastic parameters of
the individual components. According to the thermodynamical analysis
\cite{kozlov3}, the dependence of the elastic
parameters of a monolayer on its composition can have a complicated
character. In particular, it has been demonstrated for monolayers of mixtures
of diblock copolymers consisting of molecules with different block lengths 
\cite{safran5}. In a simpler model 
the spontaneous and intrinsic curvatures
are supposed to be averages of the characteristics of the components. For
example, for a two-component monolayer consisting of amphiphiles with
intrinsic curvatures $c_{in}^1$, $c_{in}^2$, the resulting spontaneous
curvature is modeled as

\begin{equation}
c_{in} = \theta c_{in}^1 + (1 - \theta) c_{in}^2
  \label{3}
\end{equation}
where $\theta$ is the molar  fraction of the first component in the
mixed system (varying from 0 to 1).

Only recently \cite{rand1,leikin} the experimental studies
allowed direct verification of this model.
Measurements of structural parameters of H$_{II}$ phases composed of two
phospholipids DOPE and dioleoylphosphatidylcholine (DOPC) \cite{rand1}
indicate that the intrinsic curvature is a
linear function of their relative composition.
Further confirmation of the linear relation (3)
over a wide range of compositions
is found in another recent study of H$_{II}$
phases consisting of DOPE and dioleoylphosphatidylglycerol (DOG)
\cite{leikin}.

The situation is less clear
for the bending modulus of mixed amphiphilic membranes.
Models of monolayers of diblock copolymers predict complex
nonlinear dependence on the composition \cite{safran5}. Similar results
follow 
from the mean-field approach \cite{benshaul}.
Furthermore, the experimental
determination of the bending elasticity of monolayers composed of DOPE and
DOG \cite{leikin} demonstrates nonlinear dependence of $\kappa$ on the
composition.
Therefore,
modeling of the bending elasticity of mixed membranes consisting of lipids
and detergents still poses a
challenge and more refined theories are needed.

Another related issue concerns the value of
the Gaussian curvature of mixed amphiphilic membranes
and
has been addressed only in few studies. Mean-field theories
\cite{benshaul,benshaul2}
computed numerically
the change of $\bar\kappa$ as function of the ratio of two
amphiphiles having different length of hydrophobic chain. 
Analytical expression for $\bar\kappa$ has been derived only for mixed
monolayers of diblock copolymers \cite{safran5}.
However,
we do not know of any
experimental investigations which directly
attempted to obtain  $\bar\kappa$ of
mixed membranes. This question remains open for both theoretical and
experimental studies.

\section{Theoretical Models of Mixed Aggregates}

Attempts have been made to interpret all types of aggregates observed in
experiments on mixed amphiphiles.
The specific questions addressed in these studies
are: (i) the existence of stable single-component
lipid vesicles characterized by a preferred size; (ii) the increase
 of vesicle sizes upon addition of small
amounts of micelle-forming surfactants;  (iii) the transition of mixed
vesicles into mixed micelles at a critical lipid-to-surfactant
ratio, and, finally, (iv) the preferred
shapes of mixed micelles.

Thermodynamic stability of single-component lipid vesicles
cannot be
understood only in terms of  model (1)
\cite{safran}. Indeed, a {\em symmetric} lipid bilayer
has a
vanishing spontaneous curvature. Therefore, the bending
energy of a spherical vesicle is equal to
$F=4\pi (2\kappa + \bar\kappa)$ (neglecting corrections
of order $\delta/R$, $\delta$ and $R$ are the
membrane thickness  and
vesicle radius,
respectively)
and does not depend on
its radius. The bending modulus
$\kappa$ is positive but the
Gaussian curvature modulus $\bar\kappa$ can have negative values.
 Depending on the relationship between $\kappa$
and $\bar\kappa$ the vesicle energy $F$ can be either
 positive or negative, and vesicles
in a suspension have, respectively, a tendency to grow
or to decrease their size.
To understand the existence of a preferable radius
of one-component vesicles, one has to account,
in addition to the bending energy (1), for other factors like
the translational entropy of vesicles
in a dilute water solution or the effects of interactions
between vesicles in dense suspensions.

Stability of mixed vesicles can be understood by
considering the possibility of a
 spontaneous curvature of bilayers
resulting from asymmetric partitioning of different amphiphiles
between the two membrane leaflets
\cite{safran}.
Analysis of the bending energy of such
systems took into account the entropy of mixing
in the membrane and an additional
interaction between the amphiphiles in the bilayer.
The model \cite{safran}
provided an explanation of experiments by
Kaler et al \cite{kaler} on formation of stable vesicles
for mixtures of cationic and anionic surfactants.

Nevertheless, the growth of vesicle size upon addition
 of micelle-forming surfactants is still
not very well understood since simple
 models predict an opposite behavior; namely, a
decrease in the vesicle size due to added surfactants. Indeed,
 the concentration of surfactant molecules should
be higher in the outer leaflet (as compared with the inner one)
of a spherical vesicle since the
curvature there has the same sign as the
spontaneous curvature of the micelle-forming surfactant.
It results in an increase of the spontaneous
curvature of the bilayer favoring formation of vesicles
of smaller radii.

The addition of
surfactant is also affecting the Gaussian curvature
modulus
of the bilayer $\bar\kappa$. Insertion of
micelle-forming molecules in membrane monolayers is known
\cite{petrov} to change $\bar\kappa$ of the
bilayer towards more negative values. It should also
 enhance the tendency of lipid vesicles to
separate into smaller aggregates rather than to grow in
size. These considerations may indicate a possible
reason for the
decrease of vesicle size as observed
very recently in systems of dimeric amphiphiles
mixed with surfactants \cite{danino}. On the other hand,
 the more common observation
of the increase in the  vesicle size
remains unexplained.

\section{Conclusions}

Theoretical descriptions of phase transition of mixed
liposomes into mixed micelles has been formulated using the
elastic model (2)-(3) \cite{andelman} and a more microscopic
mean-field approach accounting for the amphiphile
chain packing \cite{fattal}.
% The
%intrinsic curvatures are assumed (3) and equal bending moduli of
%the components.
Besides the bending energy, the entropy
of mixing of the two components
has been shown
to determine the phase behavior of the system. The
predicted phase diagram of mixed amphiphiles
has all the qualitative features of the
experimental data. An assumption of the model, based on the
 experimental observations, was that the micelles
have (in the coexistence region) the
shape of infinitely long cylinders. On the other hand,
other aggregate and micellar forms
have been suggested in the literature. The spheroidal micelles
\cite{danino} and the perforated bilayers \cite{edwards} have
 been directly observed, while the disc-like micelles
remain a hypothesis used at earlier stages to explain
 the results of dynamic light scattering. Analysis of
the factors determining the shapes of amphiphilic aggregates
resulting from solubilization of liposomes by
micelle-forming surfactants is the subject of much current interest.
Hopefully, further theory and experiments in this area will resolve
some of the outstanding issues.

\vspace{2cm}
\newlength{\tmp}
\setlength{\tmp}{\parindent}
\setlength{\parindent}{0pt}
{\em Acknowledgments}
\setlength{\parindent}{\tmp}

We benefited from discussions with M. Almgren, A. Ben-Shaul,
R. Granek, W. Helfrich, S. Leikin,
A. Parsegian, J.-L. Rigaud,
S.~Safran, Y. Talmon and R. Zana.
We are most grateful to D. Lichtenberg for introducing us
to systems of mixed amphiphiles and for many valuable discussions.
Support from the German-Israeli Foundation (G.I.F.)
under grant No.~I-0197 and the US-Israel Binational Foundation (B.S.F.)
under grant No.~94-00291 is gratefully acknowledged.

\pagebreak
%---------------------------------------------------------------
% References
%---------------------------------------------------------------

\pagebreak


\begin{thebibliography}{99}

\bibitem{luzatti} Luzatti V: {\bf The structure of the liquid
crystalline phases of lipid-water systems.} In {\sl Biological
Membranes.} Edited by Chapman D. New York: Academic Press; 1968:71-123.

\bibitem{lasic} Lasic DD: {\bf Liposomes: from physics to applications}.
Amsterdam: Elsevier Science B.V.; 1993.


\bibitem{barenholz} {\bf Handbook of non-medical
applications of liposomes}. Edited by Barenholz Y, Lasic DD. New York:
CRC Press; 1996.

\bibitem{lichtenberg} Lichtenberg D: {\bf Liposomes as a model for
solubilization and reconstitution of membranes.}
In Ref. \cite{barenholz}: 199-218. \par
$\bullet\bullet$
This is a review of the newest experimental data on structural behavior
of lipid-detergent systems with clear formulation of existing problems.

\bibitem{rand}Rand RP, Fuller N: {\bf Structural dimensions and their
changes in a re-entrant hexagonal-lamellar transition of phospholipids}.
{\sl Biophys. J.} 1994, {\bf 66}:2127-2138.

\bibitem{seddon} Seddon JM, Templer RH: {\bf Cubic phases of
self-assembled amphiphilic aggregates.} {\sl Phil. Trans. R. Soc.
Lond. A} 1993, {\bf 344}:377-401.


\bibitem{koynova} Koynova R, Caffrey M: {\bf Phases and phase transitions
of the hydrated phosphatidylethanolamine.} {\sl Chem. Phys. Lipids}
1994, {\bf 69}:1-34.

\bibitem{rigaud} Rigaud JL, Pitard B, L\'evy D:
{\bf Reconstitution of membrane proteins into liposomes:
application to energy-transducing membrane proteins}.
{\sl Biochim.  Biophys. Acta} 1995,
{\bf 1231}:223-246.


\bibitem{danino} Danino D, Talmon Y, Zana R: {\bf Vesicle to micelle
transformation in systems containing dimeric surfactants.} Submitted.
\par
{$\bullet\bullet$}
In this work a new system of consisting of dimeric surfactants is studied.
In contrast to the traditional lipid/detergent mixtures, the formation of
spheroidal micelles and
reduction in vesicle size induced by detergent
is observed.

\bibitem{edwards} Edwards K, Gustafsson J, Almgren M, Karlsson G:
{\bf Solubilization of lechitin vesicles by cationic surfactants:
intermediate structures in the vesicle-micelle transition observed
by cryo-transmission electron microscopy.} {\sl J. Coll. Interface Sci.}
1993, {\bf 161}:299-309.
\par
$\bullet\bullet$
In this work a new structure of perforated vesicles is reported. This
structure is seen as intermediate between the bilayers and the
micelles.

\bibitem{blankschtein} Zoeller NG, Blankschtein D: {\bf Development of
user-friendly computer programs to predict solution properties of single
and mixed surfactant systems.} {\sl Ind.Eng.Chem.Res.} 1995, {\bf 34}: 4150
-4160.
%\par 
%$\bullet$ In this paper gives a review of microscopic theories developed by
%the authors to describe self-assembly of surfactants.

\bibitem{helfrich} Helfrich W: {\bf Elastic properties of lipid
bilayers: theory and possible experiments.} {\sl Z. Naturforsch.}
1973, {\bf 28c}:693-703.


\bibitem{israelachvili} Israelachvili JN: {\bf Intermolecular and
surface forces} London: Academic Press; 1990.

\bibitem{gruner} Gruner SM: {\bf Intrinsic curvature hypothesis for
biomembrane lipid composition: a role for nonbilayer
lipids.}{\sl Proc. Natl. Acad. Sci. (USA)} 1985, {\bf 82}:3665-3669.


\bibitem{kozlov1} Kozlov MM, Leikin SL, Markin VS: {\bf
Elastic properties of interfaces: elastic moduli and spontaneous
geometrical characteristics.} {\sl J. Chem. Soc., Farad. Trans. 2}
1989, {\bf 85}:277-292.

\bibitem{kozlov2} Kozlov MM, Leikin SL, Rand RP: {\bf
Bending, hydration and interstitial energies quantitatively account
for the hexagonal-lamellar-hexagonal reentrant phase transition in
dioleoylphosphatidylethanolamine.} {\sl Biophys. J} 1995,
{\bf 67}:1603-1611.
\par
$\bullet$ In this work  membrane elasticity has been applied to
analyzed experimental data on reentrant
phase transitions between hexagonal and
lamellar phases.

\bibitem{benshaul} Ben-Shaul A: {\bf Molecular theory of chain packing,
elasticity and lipid-protein interaction in lipid bilayers}.
In {\sl Handbook of biological physics}. Edited by Lipowsky R. and Sackmann E.
Amsterdam: Elsevier Science B.V.; 1995.
\par
$\bullet$ This review contains an extended overview of mean-field chain packing
approach used in describing membrane elasticity and related properties.

\bibitem{kozlov3} Kozlov MM, Helfrich W: {\bf Effects of co-surfactants
on the stretching and bending elasticity of a surfactant monolayer.}
{\sl Langmuir} 1992, {\bf 8}:2792-2797.

\bibitem{safran5} Dan N., Safran SA: {\bf Self-assembly in mixtures of
diblock copolymers.} {\sl Macromolecules} 1994, {\bf 27}: 5766-5772.
\par
$\bullet$ In this work analytical expressions are derived for spontaneous
curvatures and elastic moduli of mixed monolayers of diblock copolymers
consisting on long and short blocks. In addition, a review of
related theoretical results is given.

\bibitem{rand1} Rand RP, Fuller N, Gruner SM, Parsegian VA:
{\bf Membrane curvature, lipid segregation, and structural transitions for
phospholipids under dual-solvent stress}. {\sl Biochemistry} 1990,
{\bf 29}:76-87.
\par
$\bullet$ The first work where the dependence of the spontaneous curvature
on composition has been directly measured.


\bibitem{leikin} Leikin SL, Kozlov MM, Fuller NL, Rand RP: {\bf
Measured effects of diacylglycerol on structural and elastic properties
of phospholipid membranes.} Submitted to {\sl Biophys. J.}
\par
$\bullet$
In this work the elastic properties (intrinsic curvature and bending modulus)
of two-component monolayer
are systematically studied in dependence of composition.

\bibitem{benshaul2} Szleifer I, Kramer D, Ben-Shaul A, Gelbart WM, Safran
SA: {\bf Molecular theory of curvature elasticity in surfactant films.}
{\sl Langmuir} 1990, {\bf 92}: 6800 - 6916
 

\bibitem{safran} Safran SA, Pincus P, Andelman D: {\bf Theory of
spontaneous vesicle formation in surfactant mixtures.} {\sl Science} 1990,
{\bf 248}:354-356. Safran SA, Pincus PA, Andelman D, MacKintosh FC: {\bf
Stability and phase behavior of mixed surfactant vesicles}{\sl Phys. Rev. A}
1991, {\bf 43}:1071-1078.
\par
$\bullet$
Theoretical model of stability of mixed micelles is presented in terms
asymmetric partitioning of molecules of different amphiphiles between the
outer and inner monolayers membrane. It explains the experimental results
of Kaler et. al. (Ref. \cite{kaler}).

\bibitem{kaler} Kaler EW, Murthy AK, Rodriguez BE, Zasadzinski JAN:
{\bf ...} {\sl Science} 1989, {\bf 245}:1371-xxxx.

\bibitem{petrov} Petrov AG and Bivas I: {\bf Elastic and flexoelectric
aspects of out-of-plane fluctuations in biological and model membranes.}
{\sl Prog. Surf. Sci.} 1984, {\bf 16}:389-512.

\bibitem{andelman} Andelman D, Kozlov MM, Helfrich W: {\bf Phase transition
between vesicles and micelles driven by competing curvatures}
{\sl Europhys. Lett.} 1994, {\bf 25}:231-236.
\par
$\bullet$ Theoretical model of
solubilization of liposomes by detergents is developed
and describes the
experimental phase diagrams.

\bibitem{fattal} Fattal D, Andelman D, Ben-Shaul A: {\bf The vesicle-micelle
transition in mixed lipid-surfactant systems: a molecular model.}
{\sl Langmuir} 1995, {\bf 11}:1154-1161.
\par
$\bullet$ This work complements Ref. \cite{andelman} by chain packing
mean-field 
calculations and arrives to similar conclusions.


\end{thebibliography}
\end{document}